# Tuning Structures and Electronic Spectra of Graphene Layers by Tilt Grain Boundaries

Long-Jing Yin[1,§], Jia-Bin Qiao[1,§], Wen-Xiao Wang[1], Zhao-Dong Chu[1], Kai Fen Zhang[2], Rui-Fen Dou[1,]*, Chun Lei Gao[2], Jin-Feng Jia[2], Jia-Cai Nie[1], and Lin He[1,]*

[1] Department of Physics, Beijing Normal University, Beijing, 100875, People's Republic of China

[2] Department of Physics and Astronomy, Key Laboratory of Artificial Structures and Quantum Control (Ministry of Education), Shanghai Jiao Tong University, Shanghai 200240, People's Republic of China

**Despite the structures and properties of tilt grain boundaries of graphite surface and graphene have been extensively studied, their effect on the structures and electronic spectra of graphene layers has not been fully addressed. Here we study effects of one-dimensional tilt grain boundaries on structures and electronic spectra of graphene multilayers by scanning tunneling microscopy and spectroscopy. A tilt grain boundary of a top graphene sheet in graphene multilayers leads to a twist between consecutive layers and generates superstructures (Moiré patterns) on one side of the boundary. Our results demonstrate that the twisting changes the electronic spectra of Bernal graphene bilayer and graphene trilayers dramatically. We also study quantum-confined twisted graphene bilayer generated between two adjacent tilt grain boundaries and find that the band structure of such a system is still valid even when the number of superstructures is reduced to two in one direction. It implies that the electronic structure of this system is driven by the physics of a single Moiré spot.**

More than 20 years ago, tilt grain boundaries of graphite surfaces had first been observed in scanning tunneling microscopy (STM) studies [1], and since then structures and properties of grain boundaries on graphite and graphene, a one-atom-thick hexagonal crystal of carbon atoms [2,3], have attracted much attention [4-14]. It was demonstrated that the tilt grain boundaries could show distinct electronic, magnetic, and mechanical properties depending strongly on their atomic structures [5,6,13-16]. Importantly, tilt grain boundary on graphene is more than just curiosity itself, since that this line defect can divide and disrupt the crystal of graphene and, consequently, it is expected to affect the properties of the whole system. This is especially true in graphene bilayers. A tilt grain boundary of a top graphene sheet in graphene bilayers generates Moiré patterns on one side of the boundary. The resulting stacking misorientation between adjacent layers could change the electronic properties of graphene bilayers [17-40].

Recently, several experiments demonstrated that the slightly twisted graphene bilayers exhibit low-energy linear dispersion and two low-energy saddle points in the band structure [25,30-37], as predicted by the continuum model [17]. Very recently, it was predicted that the low-energy physics in this system is still well described by the continuum model even in a single Moiré spot, which suggests that each Moiré spot in twisted graphene bilayer can be treated as a "Moiré quantum well" [41]. Despite the enormous interest in the properties of twisted graphene bilayers and the potential applications based on nanostructures of twisted graphene bilayers, the study of quantum-confined twisted graphene bilayers remains an experimental challenge. In this Letter, we study effects of one-dimensional (1D) tilt grain boundaries on structures and electronic spectra of graphene multilayers by scanning tunneling microscopy (STM) and spectroscopy (STS). We address the electronic structures of quantum-confined twisted graphene bilayers generated by two adjacent 1D tilt grain boundaries. The characteristic low-energy density of states (DOS) of twisted graphene bilayers are clearly observed even when the number of Moiré spots is reduced to two in the perpendicular direction of the grain boundaries. This directly validates the picture of Moiré quantum well in twisted graphene bilayers [41].

Our experiments were performed on highly oriented pyrolytic graphite (HOPG) surface. The HOPG samples were of ZYA grade (NT-MDT) and were surface cleaved with adhesive tape prior to experiments (See Supplementary Information [42] for details of method). A tilt grain boundary of a top graphene sheet in Bernal graphene bilayer leads to a twist between the graphene sheets and results in the emergence of Moiré patterns on one side of the boundary, as illustrated in Fig. 1a. The twisted angle between the two rotated graphene grains, $\theta$, is related to the period of the Moiré patterns by $D = a/[2\sin(\theta/2)]$ with $a \sim 0.246$ nm [17-23]. The periodicity of the one-dimensional superlattice along the grain boundary, $P_B$, is determined by the twisted angle $\theta$ and the orientation of the boundary in respect to the graphene lattice, $\alpha$ [5,7]. For $\alpha = \pm\theta/2$, $P_B = \sqrt{3}\, D$ (see Figure S1 in Supplementary Information [42] for its schematic structure); for $\alpha = 30^\circ \pm \theta/2$, we have $P_B = D$. In our experiments, the latter relation has been found to hold for all the observed grain boundaries of graphite surface, as shown in STM images of Fig. 1b-e (see Figure S2 in

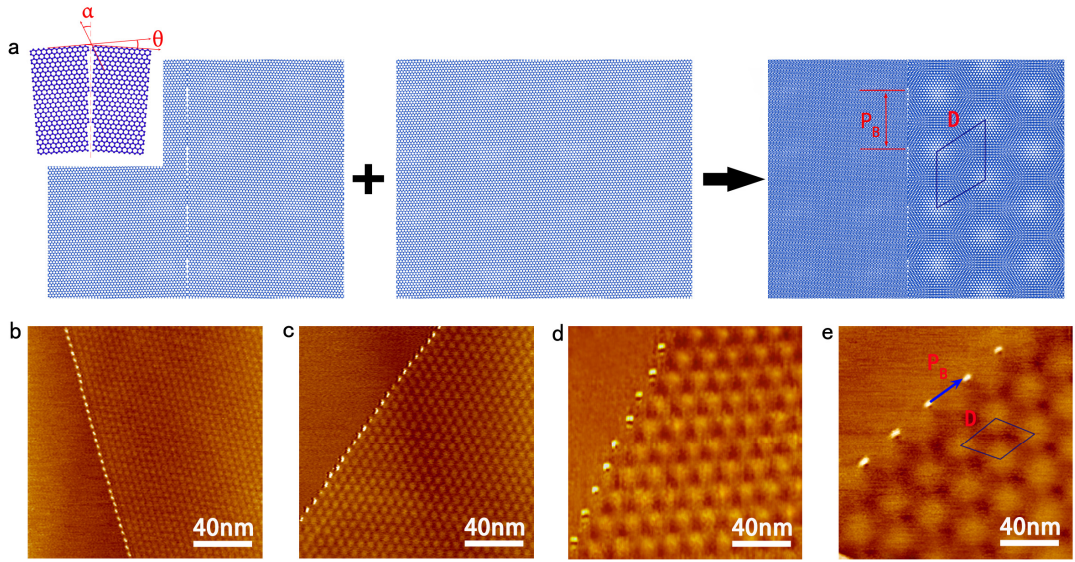

**Figure 1** (color online). Structures of graphene layers with a tilt grain boundary on the first layer. **a**, Two graphene grains meet with a relative misorientation of $\theta$ forming a tilt grain boundary. A graphene monolayer with a tilt grain boundary placed on top of another "perfect" graphene sheet forms a unique graphene bilayer, in which one side of the grain boundary is Bernal graphene bilayer and the opposite side is twisted graphene bilayer with period of Moiré pattern $D$. There are two possible tilt grain boundaries, which show different period of superstructures $P_B$ along the boundary, depending on the twisted angle $\theta$ and $\alpha$, the orientation of the boundary in respect to the graphene lattice. In panel **a**, $\alpha = 30° \pm \theta/2$, $P_B = D$. **b-e,** Large-area STM images of graphene layers on graphite surface with a tilt grain boundary on the top layer. Moiré pattern with different periods $D$ appears on one side of the tilt grain boundary. All the grain boundaries show 1D superlattices with periodicities $P_B \approx D$. **b**, $D = 4.7$ nm, $\theta = 3.0°$ ($V_{sample} = 30$ mV and $I = 1$ nA); **c**, $D = 7.2$ nm, $\theta = 2.0°$ ($V_{sample} = 50$ mV and $I = 1$ nA); **d**, $D = 17.0$ nm, $\theta = 0.83°$ ($V_{sample} = 50$ mV and $I = 1.2$ nA); **e**, $D = 28.2$ nm, $\theta = 0.50°$ ($V_{sample} = 70$ mV and $I = 1.1$ nA).

Supplementary Information [42] for more tilt grain boundaries and the resulting Moiré superstructures observed in our experiments).

Figure 2a shows the twisted angle dependence of $P_B$ for $0.37° \leq \theta \leq 3°$. It follows a simple formula $P_B = a/\theta$, which confirms the $P_B = D$ relation explicitly (for small twisted angle, $\sin\theta \sim \theta$, therefore, $D = a/\theta$). This experimental result indicates that the structure of tilt grain boundaries with $\alpha = 30° \pm \theta/2$ may be more energetic favourable than that with $\alpha = \pm\theta/2$, at least, for small twisted angles. The tilt grain boundary not only affects the structure of graphene layers, but also changes the electronic properties of the system dramatically. The system on one side of the grain boundary is changed from Bernal graphene bilayer to twisted graphene bilayer. Simultaneously, the twisting splits the parabolic spectrum of Bernal graphene bilayer into two Dirac cones. For a twisted graphene bilayer, the relative shift of the Dirac points on the different layers in the momentum space is $|\Delta K| = 2|K|\sin(\theta/2)$, where $K$ is the reciprocal-lattice vector [17-23]. Consequently, two saddle points along the intersections of two Dirac cones appear in the low-energy band spectrum. The saddle points result in two van Hove singularities (VHSs) in the density of states at energies about $E_{VHS}^{\pm} = \pm(\hbar v_F \Delta K/2 - t_\theta)$. Here $v_F$ is the Fermi velocity of the graphene, $t_\theta$ is the interlayer hopping parameter (see Supplementary Information [42] for details of analysis). Our scanning tunneling spectroscopy (STS) measurements on Moiré pattern of graphite surface, as shown in Fig. S3 of Supplementary Information [42], observe two pronounced peaks in the density of states, which should be attributed to the VHSs of twisted graphene bilayer. The energy difference of the two VHSs $\Delta E_{VHS}$ is

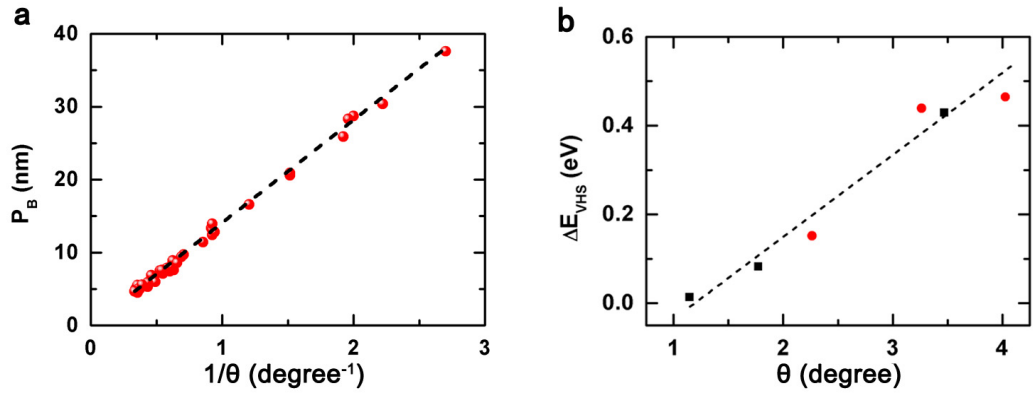

**Figure 2** (color online). **a**, The period of one-dimensional superlattice along tilt grain boundaries as a function of $1/\theta$. $\theta$ is the twisted angle of the two graphene grains connected by a tilt boundary. The solid red circles are the average experimentally measured values obtained from several tens of samples. The black dashed line is plotted according $P_B = a/\theta$ with $a$ = 0.246 nm. **b**, The energy difference of the two VHSs as a function of the twisted angles. The red solid circles are the average experimentally measured values obtained in twisted graphene bilayer on graphite surface. The black solid squares are the experimental results taken from the CVD-grown graphene sheet deposited on graphite, as reported in ref. 25. The dashed line is guide to eyes.

determined by the twisted angle and the interlayer coupling strength. The twisting-angle dependence of $\Delta E_{VHS}$, as shown in Fig. 2b, is its unmistakable signature [25,30,36,37] and indicates that the interlayer hopping parameter $t_\theta \sim 150$ meV. Additionally, the experimental observation of electronic spectra of twisted graphene bilayer on graphite surface also suggests decoupling of the surface layer from the bulk HOPG samples (or from the graphene multilayers), which consists well with the results reported in previous work [43-47].

Figure 3a shows a STM image of graphene trilayer on a graphite surface. The graphene trilayer has two adjacent tilt grain boundaries: the left boundary is on the top graphene layer, the right one is on the second layer (see Fig. 3b, Fig. 3c and Fig. S4 of Supplementary Information [42] for details of its structure). Figure 3d and 3e show the schematic structure of the graphene trilayer. The Moiré patterns in the left region of higher contrast (region **II**) arise from a stacking misorientation (with the twisted angle $\theta$) between the top graphene layer and the underlying layer. The superstructures in the right region of lower contrast (region **III**) are Moiré pattern due to a misorientation angle $\theta'$ between the second layer and the third layer. Interestingly, we have $\theta \approx \theta' \sim 2.2°$ for such a unique structure. Therefore, the top graphene layer and second layer are *AB* (Bernal) stacking in both the region **I** and **III**. The Moiré pattern due to the stacking misorientation between the first layer and the second layer is confined in region **II**. Similar structures are usually observed on graphite surfaces (see Fig. S5 and Fig. S6 in Supplementary Information [42] for more graphene trilayers with similar structures).

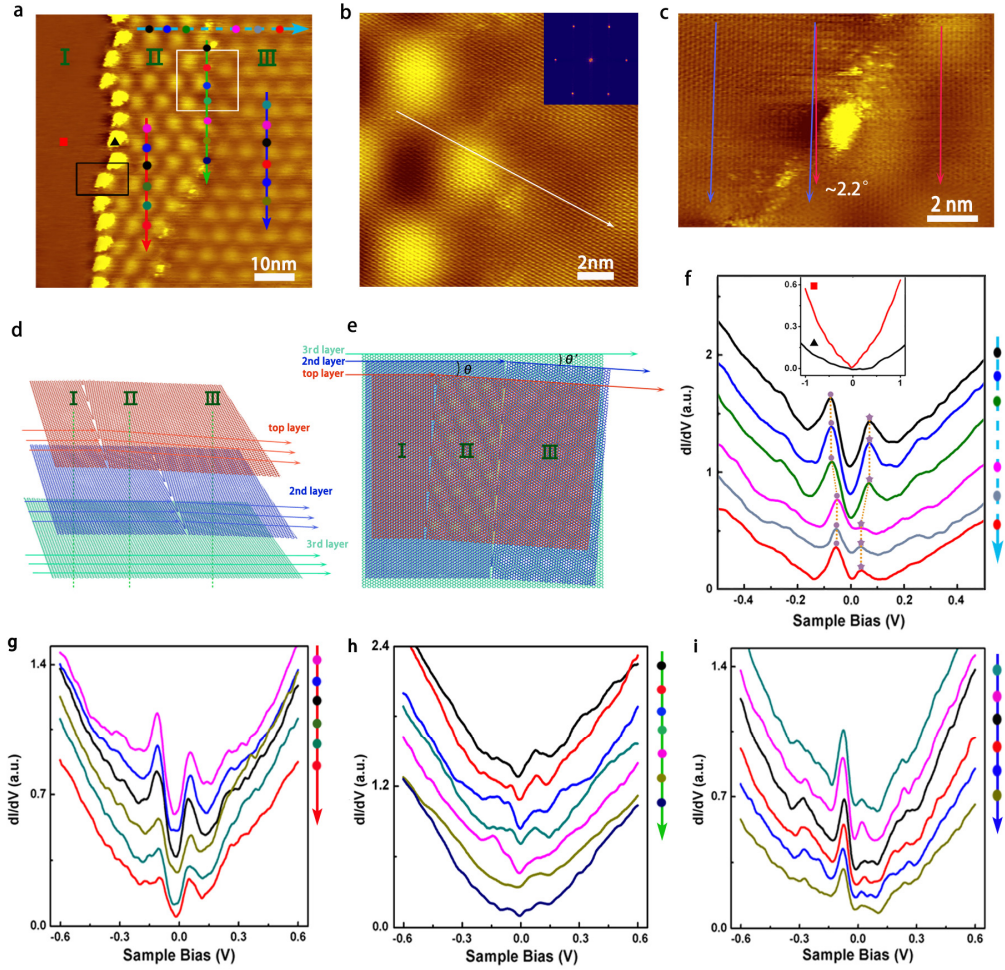

**Figure 3** (color online). Structures and electronic spectra of graphene trilayers with tilt grain boundaries on the first and second layers. **a,** Large-area STM image of graphene layers on graphite surface with two sets of superlattices showing a region of higher contrast (left) and a region of lower contrast (right) ($V_{sample}$ = 570 mV and $I$ = 44.9 pA). The left quasi-one-dimensional periodic protuberance is a tilt grain boundary of the top layer, the right one is a tilt boundary of the second layer. **b,** Zoom-in topography of the white frame in (a) with atomic resolution ($V_{sample}$ = 107 mV and $I$ = 45.4 pA). The white line shows no atomic mismatch between the two regions of superlattice, indicating that the surface graphene layer is continuous for both regions. The inset shows Fourier transforms of panel (b) with outer hexagonal spots corresponding to the atomic lattice. **c,** Zoom-in topography of the black frame in (a) with atomic resolution ($V_{sample}$ = 234 mV and $I$ = 44.4 pA). The blue and red lines give one direction of the atomic lattice of graphene grains connected by the left tilt boundary. The angle between these directions is measured to be ~ 2.2°, which agrees well with that estimated from the period of Moiré pattern. **d and e,** Schematic pictures of the structure in panel (a). The superstructures in the region of higher contrast are attributed to Moiré pattern arising from a stacking misorientation between the top graphene layer and the underlying layer. The superstructures of lower contrast are Moiré pattern due to a misorientation angle between the second layer and the third layer. We have $\theta \approx \theta'$ in the structure of panel (a). **f,** STS measurements at different positions indicated in panel (a). **g-i**, d$I$/d$V$-$V$ curves measured on different positions of panel **a**. The spectra have been vertically offset for clarity.

The unique system shown in Fig. 3a may provide a platform for exploring the finite size effect in twisted graphene bilayer [41] and for studying the effect of the third layer on the electronic band structure of twisted graphene bilayer. While the STS in the region **I** shows no discernible structure, two VHSs with energy difference of about 160 meV are clearly resolved in the spectrum measured in the region **II**, as shown in Fig. 3. In literature, the electronic structure of twisted graphene bilayer is initially calculated for periodic Moiré patterns [17]. Our experimental result indicates that such a band structure is still valid even when the superstructures are confined to several periods in one direction. Although the intensity of the two VHSs is lowered with reducing the period of the superstructures, as shown in Fig. 3g, the main feature of the two peaks in DOS (i.e., the characteristic low-energy DOS of periodic twisted graphene bilayer) is clearly reserved with the number of superstructures reducing to two in the perpendicular direction of the tilt boundary. This result implies that each Moiré spot in twisted graphene bilayer can be treated as a Moiré quantum well trapping low-energy electrons of this system [41]. It also, to some extent, demonstrates that the Dirac fermions can be localized by the Moiré pattern in twisted graphene bilayer [18,20].

In the region **III** of Fig. 3a, the stacking fault of $\theta' \approx \theta \approx 2.2°$ between the third layer and the second layer results in the Moiré patterns of almost identical period as that in the region **II**. The *AB* stacking between the top graphene layer and second layer leads to the lower contrast of the Moiré patterns (see Fig. S4 of Supplementary Information [42] for more experimental evidences). Two VHSs are also observed in

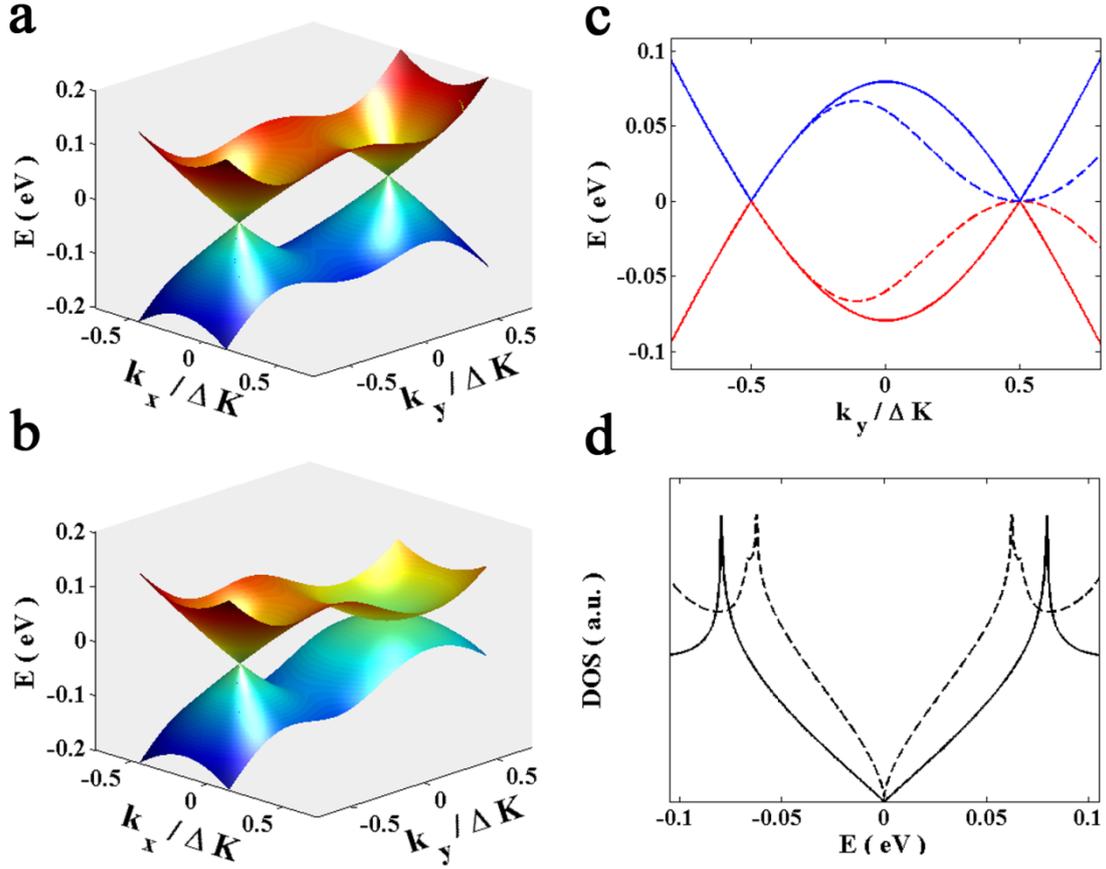

**Figure 4** (color online). Electronic spectra and low-energy density of states (DOS) of twisted graphene bilayer and trilayers. **a,** Dispersion of low-energy states for $\theta = 2.2°$, $t_\theta = 156$ meV, in a twisted graphene bilayer. Two saddle points form at $k_y = 0$ between the two Dirac cones. **b,** Electronic spectra of low-energy states for $\theta = 2.2°$, $t_\theta = 156$ meV, in a twisted graphene trilayer (the top layer and the second layer is *AB* stacking and there is a stacking fault with $\theta \approx 2.2°$ between the third layer and the second layer). Two saddle points also form in the band structure. **c,** The energy spectra are a section view of band structures with $k_x = 0$ in panel **a** (solid curve) and **b** (dashed curve). **d,** Density of states (DOS) of the bilayer in panel **a** (solid curve) and the trilayer in panel **b** (dashed curve) with VHSs (two peaks) corresponding to the energy of the two saddle points. Both are calculated numerically according to the well-known formula $\frac{S}{4\pi^2} \oint \frac{1}{\left|\nabla_k E(\vec{k})\right|} dk$. The energy difference of the two VHSs is $\Delta E_{VHS} \approx 160$ meV in the twisted graphene bilayer and is $\Delta E_{VHS} \approx 120$ meV in the twisted graphene trilayer.

the STS measurements, as shown in Fig. 3f and Fig. 3i. However, the energy difference of the two VHSs is reduced to $\Delta E_{VHS} \approx 100$ meV and the asymmetry between the intensity of the two VHSs is much strengthened. For a given twisted angle, the enhanced interlayer coupling strength will lower the energy difference of the two VHSs according to $E_{VHS}^{\pm} = \pm(\hbar v_F \Delta K/2 - t_\theta)$. Here we show that the value of $\Delta E_{VHS}$ can also be reduced significantly by introducing a third layer on top of a twisted graphene bilayer even when the interlayer coupling strength is the same. Figure 4 shows electronic spectra and low-energy density of states of twisted graphene bilayer and trilayer (Here we use the continuum model of Ref. [17]. Please see Supplementary Information [42] for details of analysis). With identical twisted angle $\theta \approx 2.2°$ and interlayer coupling strength $t_\theta \sim 156$ meV, the value of $\Delta E_{VHS}$ is about 160 meV in bilayer and is reduced to $\Delta E_{VHS} \approx 120$ meV in the trilayer.

The asymmetry between the intensity of the two VHSs, as shown in Fig. 3, was also observed in previous STS studies [25,30,31,36,37] and the possible origin of this asymmetry observed in experiments was carefully discussed in ref. 25. In the twisted graphene bilayer with a finite interlayer bias, the substrate can break the symmetry of the bilayer and generate an interlayer bias. Then the energy states at the two saddle points have different weights in the two layers. Additionally, the STM probes predominantly the local density of states of the top layer. Therefore, we expect a slight asymmetry between the intensity of the positive and negative VHSs. In the twisted graphene trilayer, the asymmetry is expected to be enhanced bacause of two main factors: first, the positive and negative states of the two saddle points will have more different weights in the underlying third layer and the top *AB* stacking bilayer; second, the STM can only probe a very weak signal of the underlying third layer. Therefore, it is reasonable to attribute the reduced $\Delta E_{VHS}$ and the enhanced asymmetry of the two VHSs in region **III** of Fig. 3a to the effect of the top graphene layer, which is *AB*-stacked with the second layer.

In summary, the experiments described here demonstrate that tilt grain boundaries can significantly affect the structures and properties of graphene multilayers. In graphene bilayer with a tilt boundary on the top layer, such a system provides an ideal platform to explore the evolution between massless Dirac fermions and massive chiral fermions around the tilt grain boundary. In graphene trilayer, tilt grain boundaries can result in the coexistence of massless Dirac fermions and massive chiral fermions of the system.

**Acknowledgements**

This work was supported by the Ministry of Science and Technology of China (Grants Nos. 2014CB920903, 2013CBA01603, 2013CB921701), the National Natural Science Foundation of China (Grant Nos. 11374035, 11004010, 51172029, 91121012), the program for New Century Excellent Talents in University of the Ministry of Education of China (Grant No. NCET-13-0054), and Beijing Higher Education Young Elite Teacher Project (Grant No. YETP0238).

§These authors contributed equally to this paper.

Email: Corresponce to Lin He (helin@bnu.edu.cn) or Rui-Fen Dou (rfdou@bnu.edu.cn).